\documentclass[letterpaper,aps,prd,superscriptaddress,showpacs,nofootinbib,floatfix,twocolumn]{revtex4}
\usepackage{graphicx}	% Include figure filed

\usepackage{xspace}	% Include xspace

\newcommand{\meanpt}{\mbox{$\langle p_T \rangle$}\xspace}

\begin{document}


\title{Cross section and double helicity asymmetry for $\eta$ mesons
and their comparison to $\pi^0$ production in $p+p$
collisions at $\sqrt{s}=200$~GeV}

\newcommand{\abilene}{Abilene Christian University, Abilene, Texas 79699, USA}
\newcommand{\acadsin}{Institute of Physics, Academia Sinica, Taipei 11529, Taiwan}
\newcommand{\banaras}{Department of Physics, Banaras Hindu University, Varanasi 221005, India}
\newcommand{\barc}{Bhabha Atomic Research Centre, Bombay 400 085, India}
\newcommand{\bnlcoll}{Collider-Accelerator Department, Brookhaven National Laboratory, Upton, New York 11973-5000, USA}
\newcommand{\bnlphys}{Physics Department, Brookhaven National Laboratory, Upton, New York 11973-5000, USA}
\newcommand{\caucr}{University of California - Riverside, Riverside, California 92521, USA}
\newcommand{\charlesczech}{Charles University, Ovocn\'{y} trh 5, Praha 1, 116 36, Prague, Czech Republic}
\newcommand{\ciae}{China Institute of Atomic Energy (CIAE), Beijing, People's Republic of China}
\newcommand{\cns}{Center for Nuclear Study, Graduate School of Science, University of Tokyo, 7-3-1 Hongo, Bunkyo, Tokyo 113-0033, Japan}
\newcommand{\colorado}{University of Colorado, Boulder, Colorado 80309, USA}
\newcommand{\columbia}{Columbia University, New York, New York 10027 and Nevis Laboratories, Irvington, New York 10533, USA}
\newcommand{\czechtech}{Czech Technical University, Zikova 4, 166 36 Prague 6, Czech Republic}
\newcommand{\dapnia}{Dapnia, CEA Saclay, F-91191, Gif-sur-Yvette, France}
\newcommand{\debrecen}{Debrecen University, H-4010 Debrecen, Egyetem t{\'e}r 1, Hungary}
\newcommand{\elte}{ELTE, E{\"o}tv{\"o}s Lor{\'a}nd University, H - 1117 Budapest, P{\'a}zm{\'a}ny P. s. 1/A, Hungary}
\newcommand{\fit}{Florida Institute of Technology, Melbourne, Florida 32901, USA}
\newcommand{\fsu}{Florida State University, Tallahassee, Florida 32306, USA}
\newcommand{\gsu}{Georgia State University, Atlanta, Georgia 30303, USA}
\newcommand{\hiroshima}{Hiroshima University, Kagamiyama, Higashi-Hiroshima 739-8526, Japan}
\newcommand{\ihepprot}{IHEP Protvino, State Research Center of Russian Federation, Institute for High Energy Physics, Protvino, 142281, Russia}
\newcommand{\illuiuc}{University of Illinois at Urbana-Champaign, Urbana, Illinois 61801, USA}
\newcommand{\instpasczech}{Institute of Physics, Academy of Sciences of the Czech Republic, Na Slovance 2, 182 21 Prague 8, Czech Republic}
\newcommand{\isu}{Iowa State University, Ames, Iowa 50011, USA}
\newcommand{\jinrdubna}{Joint Institute for Nuclear Research, 141980 Dubna, Moscow Region, Russia}
\newcommand{\kek}{KEK, High Energy Accelerator Research Organization, Tsukuba, Ibaraki 305-0801, Japan}
\newcommand{\kfki}{KFKI Research Institute for Particle and Nuclear Physics of the Hungarian Academy of Sciences (MTA KFKI RMKI), H-1525 Budapest 114, POBox 49, Budapest, Hungary}
\newcommand{\korea}{Korea University, Seoul, 136-701, Korea}
\newcommand{\kurchatov}{Russian Research Center ``Kurchatov Institute", Moscow, Russia}
\newcommand{\kyoto}{Kyoto University, Kyoto 606-8502, Japan}
\newcommand{\labllr}{Laboratoire Leprince-Ringuet, Ecole Polytechnique, CNRS-IN2P3, Route de Saclay, F-91128, Palaiseau, France}
\newcommand{\lawllnl}{Lawrence Livermore National Laboratory, Livermore, California 94550, USA}
\newcommand{\losalamos}{Los Alamos National Laboratory, Los Alamos, New Mexico 87545, USA}
\newcommand{\lpc}{LPC, Universit{\'e} Blaise Pascal, CNRS-IN2P3, Clermont-Fd, 63177 Aubiere Cedex, France}
\newcommand{\lund}{Department of Physics, Lund University, Box 118, SE-221 00 Lund, Sweden}
\newcommand{\mass}{Department of Physics, University of Massachusetts, Amherst, Massachusetts 01003-9337, USA }
\newcommand{\muenster}{Institut f\"ur Kernphysik, University of Muenster, D-48149 Muenster, Germany}
\newcommand{\muhlenberg}{Muhlenberg College, Allentown, Pennsylvania 18104-5586, USA}
\newcommand{\myongji}{Myongji University, Yongin, Kyonggido 449-728, Korea}
\newcommand{\nagasaki}{Nagasaki Institute of Applied Science, Nagasaki-shi, Nagasaki 851-0193, Japan}
\newcommand{\newmex}{University of New Mexico, Albuquerque, New Mexico 87131, USA }
\newcommand{\nmsu}{New Mexico State University, Las Cruces, New Mexico 88003, USA}
\newcommand{\ornl}{Oak Ridge National Laboratory, Oak Ridge, Tennessee 37831, USA}
\newcommand{\orsay}{IPN-Orsay, Universite Paris Sud, CNRS-IN2P3, BP1, F-91406, Orsay, France}
\newcommand{\peking}{Peking University, Beijing, People's Republic of China}
\newcommand{\pnpi}{PNPI, Petersburg Nuclear Physics Institute, Gatchina, Leningrad region, 188300, Russia}
\newcommand{\riken}{RIKEN Nishina Center for Accelerator-Based Science, Wako, Saitama 351-0198, Japan}
\newcommand{\rikjrbrc}{RIKEN BNL Research Center, Brookhaven National Laboratory, Upton, New York 11973-5000, USA}
\newcommand{\rikkyo}{Physics Department, Rikkyo University, 3-34-1 Nishi-Ikebukuro, Toshima, Tokyo 171-8501, Japan}
\newcommand{\saispbstu}{Saint Petersburg State Polytechnic University, St. Petersburg, Russia}
\newcommand{\saopaulo}{Universidade de S{\~a}o Paulo, Instituto de F\'{\i}sica, Caixa Postal 66318, S{\~a}o Paulo CEP05315-970, Brazil}
\newcommand{\seoulnat}{Seoul National University, Seoul, Korea}
\newcommand{\stonybrkc}{Chemistry Department, Stony Brook University, SUNY, Stony Brook, New York 11794-3400, USA}
\newcommand{\stonycrkp}{Department of Physics and Astronomy, Stony Brook University, SUNY, Stony Brook, New York 11794-3400, USA}
\newcommand{\subatech}{SUBATECH (Ecole des Mines de Nantes, CNRS-IN2P3, Universit{\'e} de Nantes) BP 20722 - 44307, Nantes, France}
\newcommand{\tenn}{University of Tennessee, Knoxville, Tennessee 37996, USA}
\newcommand{\titech}{Department of Physics, Tokyo Institute of Technology, Oh-okayama, Meguro, Tokyo 152-8551, Japan}
\newcommand{\tsukuba}{Institute of Physics, University of Tsukuba, Tsukuba, Ibaraki 305, Japan}
\newcommand{\vandy}{Vanderbilt University, Nashville, Tennessee 37235, USA}
\newcommand{\waseda}{Waseda University, Advanced Research Institute for Science and Engineering, 17 Kikui-cho, Shinjuku-ku, Tokyo 162-0044, Japan}
\newcommand{\weizmann}{Weizmann Institute, Rehovot 76100, Israel}
\newcommand{\yonsei}{Yonsei University, IPAP, Seoul 120-749, Korea}
\affiliation{\abilene}
\affiliation{\acadsin}
\affiliation{\banaras}
\affiliation{\barc}
\affiliation{\bnlcoll}
\affiliation{\bnlphys}
\affiliation{\caucr}
\affiliation{\charlesczech}
\affiliation{\ciae}
\affiliation{\cns}
\affiliation{\colorado}
\affiliation{\columbia}
\affiliation{\czechtech}
\affiliation{\dapnia}
\affiliation{\debrecen}
\affiliation{\elte}
\affiliation{\fit}
\affiliation{\fsu}
\affiliation{\gsu}
\affiliation{\hiroshima}
\affiliation{\ihepprot}
\affiliation{\illuiuc}
\affiliation{\instpasczech}
\affiliation{\isu}
\affiliation{\jinrdubna}
\affiliation{\kek}
\affiliation{\kfki}
\affiliation{\korea}
\affiliation{\kurchatov}
\affiliation{\kyoto}
\affiliation{\labllr}
\affiliation{\lawllnl}
\affiliation{\losalamos}
\affiliation{\lpc}
\affiliation{\lund}
\affiliation{\mass}
\affiliation{\muenster}
\affiliation{\muhlenberg}
\affiliation{\myongji}
\affiliation{\nagasaki}
\affiliation{\newmex}
\affiliation{\nmsu}
\affiliation{\ornl}
\affiliation{\orsay}
\affiliation{\peking}
\affiliation{\pnpi}
\affiliation{\riken}
\affiliation{\rikjrbrc}
\affiliation{\rikkyo}
\affiliation{\saispbstu}
\affiliation{\saopaulo}
\affiliation{\seoulnat}
\affiliation{\stonybrkc}
\affiliation{\stonycrkp}
\affiliation{\subatech}
\affiliation{\tenn}
\affiliation{\titech}
\affiliation{\tsukuba}
\affiliation{\vandy}
\affiliation{\waseda}
\affiliation{\weizmann}
\affiliation{\yonsei}
\author{A.~Adare} \affiliation{\colorado}
\author{S.~Afanasiev} \affiliation{\jinrdubna}
\author{C.~Aidala} \affiliation{\columbia} \affiliation{\mass}
\author{N.N.~Ajitanand} \affiliation{\stonybrkc}
\author{Y.~Akiba} \affiliation{\riken} \affiliation{\rikjrbrc}
\author{H.~Al-Bataineh} \affiliation{\nmsu}
\author{J.~Alexander} \affiliation{\stonybrkc}
\author{K.~Aoki} \affiliation{\kyoto} \affiliation{\riken}
\author{L.~Aphecetche} \affiliation{\subatech}
\author{R.~Armendariz} \affiliation{\nmsu}
% ...
\collaboration{PHENIX Collaboration} \noaffiliation

\date{\today}

\begin{abstract}

Measurements of double helicity asymmetries in inclusive hadron 
production in polarized $p+p$ collisions are sensitive to 
helicity-dependent parton distribution functions, in particular to 
the gluon helicity distribution, $\Delta g$. This study focuses on the 
extraction of the double helicity asymmetry in $\eta$ production 
($\vec{p}+\vec{p} \rightarrow \eta+X$), the $\eta$ cross section and 
the $\eta/\pi^0$ cross section ratio. The cross section and ratio 
measurements provide essential input for the extraction of 
fragmentation functions that are needed to access the 
helicity-dependent parton distribution functions.

\end{abstract}

	
\maketitle

\section{Introduction}

Until recently, the knowledge about helicity--dependent parton 
distribution functions (PDFs) in the nucleon mainly came from 
next-to-leading order (NLO) QCD fits (see, e.g., 
\cite{Blumlein:2010rn}) to the helicity--dependent structure function 
$g_1$, as measured in fixed--target polarized inclusive 
deep--inelastic scattering (DIS) experiments (see, e.g., 
\cite{Airapetian:2007mh,Alexakhin:2006vx}). The resulting 
helicity--dependent PDF for the gluon has rather large uncertainties 
due to the fact that the exchanged virtual photon does not couple 
directly, i.e., at leading order, to the gluon, and that an indirect 
way of accessing it via NLO fits to $g_1$ suffers from the limited 
kinematic reach of the fixed target experiments. Accessing the 
helicity--dependent gluon PDF via the so-called photon-gluon-fusion 
process in semi-inclusive DIS has not yet resulted in better 
constraints, see Ref.~\cite{:2009ey,:2010um} and references therein 
for details. Thus, additional data from polarized $p+p$ scattering, in 
which longitudinally polarized gluons are directly probed via 
scattering off longitudinally polarized gluons or quarks, should 
reduce the uncertainties in the helicity--dependent gluon PDF. This 
has been demonstrated in a global NLO fit \cite{deFlorian:2008mr} 
using, for the first time, the available inclusive and semi-inclusive 
polarized DIS data together with first results from polarized $p+p$ 
scattering at the Relativistic Heavy Ion Collider (RHIC). The results 
included were the double helicity asymmetries in inclusive $\pi^0$ 
\cite{Adare:2007dg,Adare:2008px,Adare:2008qb} and jet 
\cite{Abelev:2007vt} production from the PHENIX and STAR experiments, 
respectively.

The double helicity asymmetry in inclusive hadron production is given as
\begin{eqnarray}
\label{cross_sec_asy} %\nonumber
A_{LL} & = & \frac{\sigma^{++}-\sigma^{+-}}{\sigma^{++}+\sigma^{+-}} \\
&  = & \frac{\sum_{abc}\Delta f_a\otimes
\Delta f_b\otimes\Delta\hat{\sigma}^{ab\rightarrow cX'}\otimes D^{h}_c}{2 \sigma},
\nonumber
\end{eqnarray}
where the cross section $\sigma^{++}$ ($\sigma^{+-}$) describes the 
reaction where both protons have the same (opposite) helicity. The 
helicity--independent cross section is defined as $\sigma = 
(\sigma^{++}+\sigma^{+-})/2$. The helicity--dependent decomposition of 
the numerator is given on the right-hand side of 
Eq.~\ref{cross_sec_asy}, where $\Delta f_a$, $\Delta f_b$ represent 
the helicity--dependent PDFs for quarks or gluons, and $\Delta 
\hat{\sigma}$ are the helicity--dependent hard scattering cross 
sections calculable in perturbative QCD (pQCD). The kinematic 
dependences of these terms are omitted for simplicity. At leading 
order (LO) the fragmentation functions $D^h_c$ can be interpreted as 
the probability for a certain parton $c$ to fragment into a certain 
hadron $h$ and thus they are not needed in the case of jet and direct 
photon production. In current global fits of parton helicity 
distributions, the fragmentation functions are assumed to be spin 
independent.

This study focuses on the midrapidity cross section and double 
helicity asymmetry in inclusive $\eta$ production ($\vec{p}+\vec{p} 
\rightarrow \eta+X$) as a function of transverse momentum ($p_T$) and 
the $\eta/\pi^0$ cross section ratio at $\sqrt{s} = 200$~GeV measured 
at the PHENIX experiment at RHIC. The measurement of the $\eta$ double 
helicity asymmetry adds independent data with different systematics to 
the present set of polarized data available to PDF fits. Even when 
compared to a closely related data set, e.g., the PHENIX $\pi^0$ data 
on double helicity asymmetries, the difference in the fragmentation 
functions can lead to a different sensitivity to certain 
helicity--dependent PDFs. In contrast to the $\pi^0$, the experimental 
data available to extract $\eta$ fragmentation functions is rather 
limited. The existing $\eta$ cross section measurements from $e^++e^-$ 
collider data can constrain the quark fragmentation functions to some 
degree, but the extraction of gluon fragmentation functions requires 
either rather precise $e^++e^-$ data taken in a wide range of center 
of mass energies, or cross section measurements from processes where 
gluons are directly involved, e.g., $p+p$ scattering. Therefore, the 
data on cross sections and cross section ratios presented here serve 
as important input for the extraction of fragmentation functions, in 
particular as the measurement has been performed over a wide range of 
$p_T$.

\section{Extraction of $\eta$ and $\pi^0$ yields}

The $\eta$ ($\pi^0$) meson is reconstructed through its main decay 
channel, $\eta \, (\pi^0) \rightarrow \gamma \gamma$, with a branching 
ratio (BR) of about 39\% (99\%)~\cite{Amsler:2008zzb}. The data were 
taken at the PHENIX \cite{Adcox:2003zm} experiment in 2005 and 2006. 
After data quality and vertex cuts, $2.5\,$pb$^{-1}$ from the 2005 
data and $6.5\,$pb$^{-1}$ from the 2006 data are used for the 
analysis. The data sets from both years are used for the extraction of 
the statistics--limited double helicity asymmetry in $\eta$ production 
while only the larger data set from 2006 is used for the extraction of 
the predominantly systematics--limited cross section measurements. 
Note that the analysis described in the following, including all cuts, 
is done in the exact same way for the $\eta$ and the $\pi^0$ meson in 
order to minimize the systematic uncertainties on the $\eta/\pi^0$ 
cross section ratio presented below.

Two data sets have been analyzed, collected by requiring two different 
trigger selections. The minimum bias (MB) trigger requires coincident 
signals in two beam-beam counters (BBCs)~\cite{Allen:2003zt}, which 
are arrays of quartz-radiator \v{C}erenkov counters providing full 
azimuthal coverage at pseudorapidities of $3.0 < |\eta| < 3.9$. Based 
on the timing of the signals from the two BBCs, the event vertex is 
reconstructed and required to be within 30~cm of the nominal 
interaction point. In addition to the MB trigger, the high-$p_T$ 
triggered data set requires an energy deposition larger than 
approximately 1.4~GeV in an area of $4 \times 4$ towers in the 
electromagnetic calorimeter (EMCal)~\cite{Aphecetche:2003zr}.

The EMCal is the primary detector used in this analysis, located at a 
radial distance of about 5~m from the beam pipe. It covers the 
pseudo--rapidity range $|\eta| < 0.35$ and has an azimuthal acceptance 
of $\Delta \phi = \pi$. The EMCal consists of eight sectors, six of 
which are composed of a total of 15552 lead--scintillator (PbSc) 
sandwich towers (5.5~cm x 5.5~cm x 37.5~cm), and two sectors of 
lead-glass (PbGl) \v{C}erenkov calorimeters, consisting of a total of 
9216 towers (4~cm x 4~cm x 40~cm). For the cross section measurements 
only the PbSc was used.

A cluster in the EMCal is assumed to originate from a photon if the 
following criteria are met. First, since showers in the EMCal are not 
confined to a single tower, a shower profile analysis can be used to 
reject hadrons, which usually produce broader showers than photons. 
Since hadrons are slower than photons, an additional time of flight 
cut is used for photon identification. Furthermore, the cluster must 
not be associated with a hit from a charged particle in the Pad 
Chamber (PC3) just in front of the EMCal; an exception is made if the 
hit position in the EMCal and in the PC3 are aligned in such a way 
that the particle could have come from the vertex on a straight line, 
i.e., it was not bent in the central magnetic field. In this case, the 
cluster is accepted as a photon candidate since it is likely that the 
original photon converted into an $e^+ e^-$ pair before the PC3 but 
outside the magnetic field. The latter two selection cuts are used in 
the analysis of the double helicity asymmetry but not in the 
extraction of the cross sections, leading to a smaller signal to 
background ratio in the cross section measurements. In order to 
exclude clusters with potentially incorrectly reconstructed energies 
due to leakage effects, the tower with the largest energy deposition 
in a cluster must not be in the outermost two columns or rows of an 
EMCal sector. In addition, there must not be a noisy or dead tower in 
the eight towers surrounding the central one.

Using all possible pairs of photon candidates, the two--photon 
invariant mass spectrum is calculated. An upper limit of 0.7 is placed 
on the energy asymmetry, $|E_1 - E_2|/(E_1 + E_2)$, of the two cluster 
energies, $E_1$ and $E_2$, in order to reduce the combinatorial 
background due to numerous low-energy background clusters. It is also 
checked that either of the two clusters coincides with an area in the 
EMCal that caused a high-$p_T$ trigger. Finally, the $p_T$ of the 
diphoton is required to be larger than 2~GeV/$c$. At smaller $p_T$, 
the uncertainties in the cross section extraction become too large due 
to large backgrounds and limited acceptance.

%%%%%%%%%%%%%%%%%%%%%%%%%%%%%%%%%%%%%%%%%%%%%%%%%%%%%%%%%%%%% Fig_1
\begin{figure}
\includegraphics[width=1.0\linewidth]{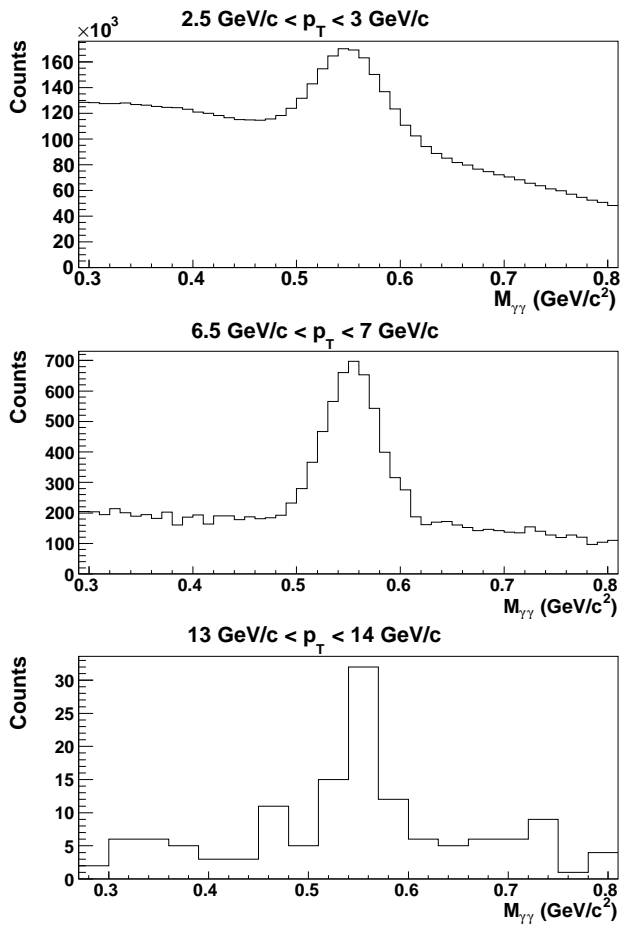}
\caption{Invariant mass distributions in the
vicinity of the $\eta$ peak for the high-$p_T$ triggered data set 
and for three different bins of $p_T$. The selection cuts for the cross section
extraction are used.}
\label{invmass}
\end{figure}

The $\eta$ and $\pi^0$ cross sections and the $\eta$ double helicity 
asymmetry are extracted in bins of $p_T$. Using the selection cuts for 
the cross section extraction on the high-$p_T$ triggered data set, the 
resulting invariant mass distributions in the vicinity of the $\eta$ 
peak are shown in Fig.~\ref{invmass} for three different bins of 
$p_T$. For bins at small $p_T$, the signal extraction is based on fits 
to the invariant mass distributions using a Gaussian for the signal 
plus a second-order polynomial for the background that describe the 
vicinity of the $\eta$ and $\pi^0$ peaks very well. For $p_T \gtrsim 
10$~GeV/$c$ the signal extraction based on fits becomes unreliable as 
limited statistics leads to large fluctuations in the fit results for 
the mean and width of the peaks. Therefore, the mean and width are 
taken from a Monte-Carlo simulation, which describes the mean and 
width of the $\eta$ and $\pi^0$ peaks as a function of $p_T$ very well 
for bins below $p_T \approx 10$~GeV/$c$, giving confidence in using 
them for the bins above. In these cases, the number of background 
counts under the signal peak is estimated by using the number of 
counts in the sidebands. The sidebands are on both sides of the mean 
of the peak, between 4 and 7 (4 and 6) times the Gaussian width of the 
peak for the cross section (double helicity asymmetry) analysis. 
However, the exact position and width of the sidebands are varied and 
possible effects are taken into account in the systematic uncertainty. 
In the mid-$p_T$ range between $3$~GeV/$c$ and $10$~GeV/$c$, where 
statistics is sufficient for the fit results to be stable and the 
background in the vicinity of the peaks is approximately linear so 
that the sideband subtraction is applicable, both methods agree as 
expected.

\section{The $\eta$ cross section and $\eta/\pi^0$ cross section ratio}

The $\eta$ and $\pi^0$ meson cross sections are calculated from
\begin{equation}
\label{cross_sec}
E \frac{d^3 \sigma} {d^3 p} = \frac{1}{2 \pi \, p_T} \frac{1}{BR} \frac{1}{\mathcal{L}}
\frac{1}{A \, \epsilon_{\rm trig} \, \epsilon_{\rm rec}} \frac{N(\Delta p_T, \Delta y)}{\Delta p_T \Delta y},
\end{equation}
where $\mathcal{L}$ denotes the integrated luminosity, $A$ the 
acceptance, $\epsilon_{\rm trig}$ the trigger efficiency, 
$\epsilon_{\rm rec}$ the reconstruction efficiency, and $N$ the 
number of reconstructed mesons.

The luminosity is calculated from the number of MB events divided by 
the cross section for events selected by the MB trigger. For the 
latter, a value of 23.0~mb with a systematic uncertainty of 9.7\% has 
been derived from Vernier scan results~\cite{Adler:2003pb} and an 
extrapolation for subsequent years. The acceptance is calculated from 
a Monte-Carlo simulation using, as an input, the map of noisy and dead 
cells also used in the data analysis. The systematic uncertainty on 
the acceptance calculation is 3.6\% (3\%) for the $\eta$ ($\pi^0$) 
meson. The trigger efficiency for the MB data is given by the MB 
trigger efficiency for $\eta$ and $\pi^0$ production. The trigger 
efficiency for the high-$p_T$ triggered data set is given by the MB 
trigger efficiency times the efficiency of the high-$p_T$ trigger. The 
MB (high--$p_T$) trigger efficiency is determined by the ratio of the 
number of reconstructed $\pi^0$ or $\eta$ mesons in a high--$p_T$ (MB) 
triggered sample in coincidence with the MB (high--$p_T$) trigger 
divided by the number of reconstructed $\pi^0$ or $\eta$ events 
without the coincidence. The MB trigger efficiency is 0.78 for both 
$\pi^0$ and $\eta$ mesons over the whole range of $p_T$ considered, 
with a systematic uncertainty of 3\%. The high--$p_T$ trigger 
efficiency reaches a plateau at a level of 0.90 for $p_T > 
4.5$~GeV/$c$ and is also very similar for both mesons. Due to the fact 
that the turn-on curve of the high $p_T$ trigger is very steep and 
reaches an efficiency of about 0.80 in the $3 < p_T < 3.5$~GeV/$c$ 
bin, the efficiency for $p_T < 3$~GeV/$c$ has a large systematic 
uncertainty. Therefore, the cross section calculation is based on the 
smaller MB triggered data set for $p_T < 3$~GeV/$c$ and on the 
high-$p_T$ triggered data set for larger transverse momenta. The 
reconstruction efficiency accounts for loss of photons due to 
conversion ($6$\% $\pm 2$\%) and due to the cut on the shower shape 
discussed above ($4$\% $\pm 2$\%). In the case of $\pi^0$ production, 
merging of the two decay photons into a single cluster is considered 
for $p_T > 10$~GeV/$c$.

%%%%%%%%%%%%%%%%%%%%%%%%%%%%%%%%%%%%%%%%%%%%%%%%%%%%%%%%%%%%% Fig_2
\begin{figure}[tb]
%\begin{center}
\includegraphics[width=1.0\linewidth] {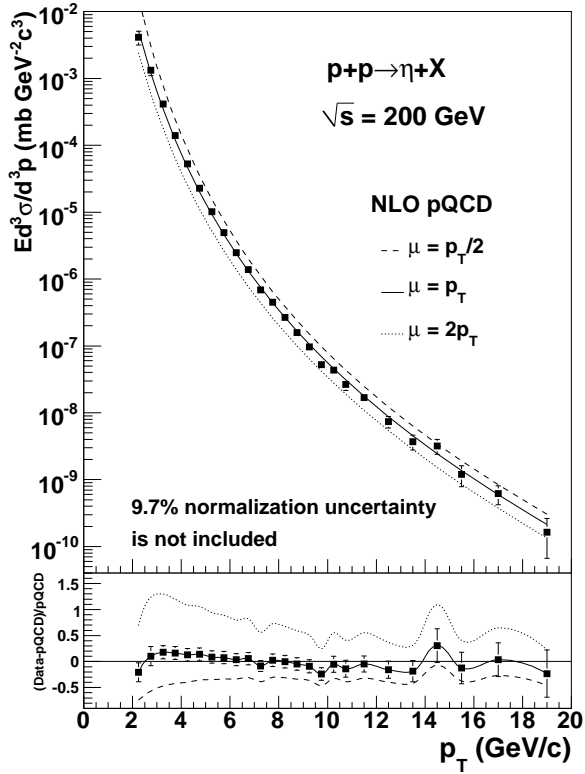}
%\end{center}
\caption{
Cross section for midrapidity inclusive $\eta$ production at
$\sqrt{s}=200$~GeV as a function of $p_T$ and its comparison to NLO 
pQCD calculations at three different scales $\mu$. The error bars 
shown are the statistical and systematic uncertainties added in 
quadrature. Not included is the overall normalization uncertainty of 
9.7\%. Note that the fragmentation functions used in the calculations 
are partially constrained by this data. See text for details.}
\label{eta_cs}
\end{figure}

The $\eta$ cross section as a function of $p_T$ between 2 and 
20~GeV/$c$ is shown in Fig.~\ref{eta_cs} and tabulated in 
Table~\ref{table:cross}. Note that a bin--shift correction is applied 
in order to be able to plot each data point at the center of each 
given $p_T$ bin, which, due to the exponentially falling spectrum, 
does not represent the true physical value of the yield in that 
bin~\cite{Lafferty:1994cj}. This in particular facilitates the 
calculation of the $\eta$/$\pi^0$ cross section ratio.

%==============================================  Table_I
\begin{table}[bt]
 \caption{\label{table:cross}
Results, statistical, and systematic (type-$B_1$, type-$B_2$) 
uncertainties of the measured $\eta$ cross sections from the 2006 data 
set. There is an additional normalization uncertainty of 9.7\% 
(type-$C$).}
 \begin{ruledtabular}
 \begin{tabular}{ccccc}
$p_T$ & $E\frac{d^3\sigma}{dp^3}$ &  Stat. unc.  & type-$B_1$ & type-$B_2$ \\
(GeV/$c$) & (mb GeV$^{-2}c^3$) &   &   &   \\\hline
2.25 & 4.12$\times 10^{-03}$ & 2.21$\times 10^{-04}$ & 8.64$\times 10^{-04}$ & 3.17$\times 10^{-04}$ \\
2.75 & 1.33$\times 10^{-03}$ & 6.98$\times 10^{-05}$ & 1.86$\times 10^{-04}$ & 1.02$\times 10^{-04}$ \\
3.25 & 4.14$\times 10^{-04}$ & 1.66$\times 10^{-06}$ & 1.65$\times 10^{-05}$ & 4.09$\times 10^{-05}$ \\
3.75 & 1.40$\times 10^{-04}$ & 7.48$\times 10^{-07}$ & 5.61$\times 10^{-06}$ & 1.39$\times 10^{-05}$ \\
4.25 & 5.28$\times 10^{-05}$ & 3.75$\times 10^{-07}$ & 2.11$\times 10^{-06}$ & 5.33$\times 10^{-06}$ \\
4.75 & 2.28$\times 10^{-05}$ & 2.12$\times 10^{-07}$ & 9.12$\times 10^{-07}$ & 2.30$\times 10^{-06}$ \\
5.25 & 1.01$\times 10^{-05}$ & 1.28$\times 10^{-07}$ & 4.05$\times 10^{-07}$ & 1.02$\times 10^{-06}$ \\
5.75 & 4.95$\times 10^{-06}$ & 8.19$\times 10^{-08}$ & 1.98$\times 10^{-07}$ & 5.00$\times 10^{-07}$ \\
6.25 & 2.48$\times 10^{-06}$ & 5.43$\times 10^{-08}$ & 9.94$\times 10^{-08}$ & 2.51$\times 10^{-07}$ \\
6.75 & 1.39$\times 10^{-06}$ & 3.70$\times 10^{-08}$ & 5.56$\times 10^{-08}$ & 1.40$\times 10^{-07}$ \\
7.25 & 6.87$\times 10^{-07}$ & 2.61$\times 10^{-08}$ & 2.75$\times 10^{-08}$ & 7.07$\times 10^{-08}$ \\
7.75 & 4.50$\times 10^{-07}$ & 1.88$\times 10^{-08}$ & 1.80$\times 10^{-08}$ & 4.63$\times 10^{-08}$ \\
8.25 & 2.67$\times 10^{-07}$ & 1.39$\times 10^{-08}$ & 1.07$\times 10^{-08}$ & 2.75$\times 10^{-08}$ \\
8.75 & 1.59$\times 10^{-07}$ & 1.04$\times 10^{-08}$ & 6.34$\times 10^{-09}$ & 1.63$\times 10^{-08}$ \\
9.25 & 9.63$\times 10^{-08}$ & 7.83$\times 10^{-09}$ & 3.85$\times 10^{-09}$ & 1.03$\times 10^{-08}$ \\
9.75 & 5.24$\times 10^{-08}$ & 5.82$\times 10^{-09}$ & 2.09$\times 10^{-09}$ & 5.60$\times 10^{-09}$ \\
10.25 & 4.33$\times 10^{-08}$ & 4.81$\times 10^{-09}$ & 1.73$\times 10^{-09}$ & 4.81$\times 10^{-09}$ \\
10.75 & 2.66$\times 10^{-08}$ & 3.94$\times 10^{-09}$ & 1.07$\times 10^{-09}$ & 2.96$\times 10^{-09}$ \\
11.5 & 1.68$\times 10^{-08}$ & 1.88$\times 10^{-09}$ & 6.74$\times 10^{-10}$ & 1.87$\times 10^{-09}$ \\
12.5 & 7.37$\times 10^{-09}$ & 1.14$\times 10^{-09}$ & 2.95$\times 10^{-10}$ & 8.55$\times 10^{-10}$ \\
13.5 & 3.70$\times 10^{-09}$ & 7.94$\times 10^{-10}$ & 1.48$\times 10^{-10}$ & 4.48$\times 10^{-10}$ \\
14.5 & 3.19$\times 10^{-09}$ & 6.70$\times 10^{-10}$ & 1.27$\times 10^{-10}$ & 4.05$\times 10^{-10}$ \\
15.5 & 1.20$\times 10^{-09}$ & 3.79$\times 10^{-10}$ & 4.78$\times 10^{-11}$ & 1.52$\times 10^{-10}$ \\
17 & 6.17$\times 10^{-10}$ & 1.74$\times 10^{-10}$ & 2.47$\times 10^{-11}$ & 8.26$\times 10^{-11}$ \\
19 & 1.64$\times 10^{-10}$ & 9.49$\times 10^{-11}$ & 6.57$\times 10^{-12}$ & 2.20$\times 10^{-11}$ \\
 \end{tabular}
 \end{ruledtabular}
 \end{table}

The $\eta$ cross section is consistent with an earlier PHENIX 
measurement \cite{Adler:2006bv} covering a smaller range in $p_T$ from 
2.5 to 12 GeV/$c$. The error bars shown in Fig.~\ref{eta_cs} are the 
statistical and systematic uncertainties added in quadrature. Not 
included is an overall normalization uncertainty of 9.7\% due to the 
uncertainty in the luminosity measurement. The other dominant 
systematic uncertainties are an approximately $p_T$-independent 
uncertainty of about 8\% due to the uncertainty on the global energy 
scale of 1.2\%, possible non--linearities in the energy scale 
affecting mainly points with $p_T > 10$~GeV/$c$, and uncertainties 
from the signal extraction affecting principally the two lowest $p_T$ 
points, which have a large background underneath the $\eta$ peak. The 
systematic uncertainties are subdivided into uncertainties that are 
uncorrelated between $p_T$ bins (type-$A$), correlated between $p_T$ 
bins (type-$B$), and overall normalization uncertainties (type-$C$). 
As described above, the peak extraction is based on different methods 
depending on $p_T$. Thus, the $p_T$ bins in certain regions are 
correlated, but there is no full correlation over the whole range. 
Such kind of uncertainties are sub-categorized as type-$B_1$, in order 
to distinguish from those correlated over all $p_T$ bins (type-$B_2$). 
All other uncertainties, except the one from the luminosity 
measurement (type-$C$), are assumed to be in this category.

The $\eta$ cross section from $p+p$ scattering presented here, 
together with the above mentioned earlier PHENIX data in a smaller 
range in $p_T$, and various $\eta$ cross section measurements from 
$e^++e^-$ scattering have been used in a global fit to extract new 
fragmentation functions for $\eta$ production at NLO~\cite{Aidala:2010bn}. 
Earlier determinations of $\eta$ fragmentation functions based on 
SU(3) model estimates at LO and normalizations taken from a Monte 
Carlo event generator at NLO are described in 
Refs.~\cite{Indumathi:1998am,Indumathi:2008mu} and 
Ref.~\cite{Greco:1993tt}, respectively. Due to the absence of data on 
semi-inclusive $\eta$ production the fragmentation functions can only 
be extracted separately for each quark flavor with additional 
assumptions. The assumption that all light quark fragmentation 
functions are the same, i.e., $D_u^\eta = D_d^\eta = D_s^\eta = 
D_{\bar u}^\eta = D_{\bar d}^\eta = D_{\bar s}^\eta$, has been used in 
Ref.~\cite{Aidala:2010bn}. Using these fragmentation functions and the 
CTEQ6M~\cite{Pumplin:2002vw} PDFs as an input to the NLO code of 
Ref.~\cite{Jager:2002xm}, pQCD calculations at three different scales 
$\mu$ are carried out. Here, $\mu$ represents the factorization, 
renormalization and fragmentation scales, i.e., the three scales are 
set equal in each separate calculation. With these new fragmentation 
functions, for which the present data constitute nearly 20\% of the 
input experimental data points, the cross section is described well.

\begin{figure}[t] 
\includegraphics[width=1.0\linewidth]{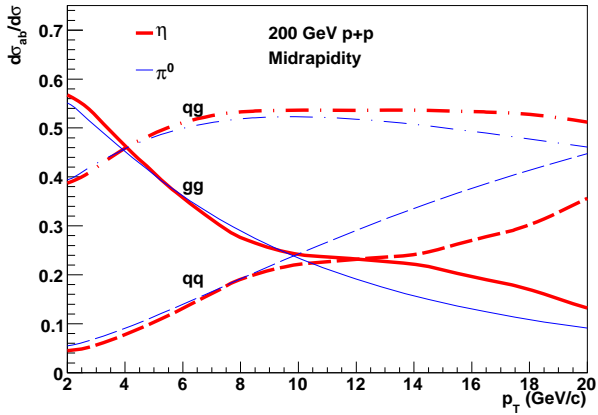} 
\caption{(color online) 
Fractional contribution of gluon--gluon ($gg$), 
quark--gluon ($qg$), and quark--quark ($qq$) scattering to the $\eta$ 
production in the pQCD calculation of Fig.~\ref{eta_cs}, and to the 
$\pi^0$ production~\cite{deFlorian:2007aj}, as a function of $p_T$.}
\label{subprocess_frac}
\end{figure}

The contributions of the various scattering subprocesses, gluon--gluon 
($gg$), quark--gluon ($qg$), and quark--quark ($qq$), to the $\eta$ 
production as a function of $p_T$, are shown in 
Fig.~\ref{subprocess_frac}. For comparison, they are also shown in the 
case of $\pi^0$ production~\cite{deFlorian:2007aj}. While the 
corresponding uncertainties are difficult to quantify, it is clear 
that the subprocess contributions to the $\eta$ and $\pi^0$ production 
are, within uncertainties, identical up to a $p_T$ of approximately 
10~GeV/$c$. This is the kinematic range of the $\eta$ and $\pi^0$ 
double helicity asymmetries presented below and published in 
Ref.~\cite{Adare:2007dg,Adare:2008px}, respectively. Consequently, 
these measurements have approximately the same sensitivity to the 
gluon helicity distribution accessible via the $gg$ and $qg$ 
subprocesses. The differences at larger values of $p_T$ are mostly 
related to uncertainties in the fragmentation functions and thus do 
not necessarily indicate different sensitivities to polarized PDFs.

%%%%%%%%%%%%%%%%%%%%%%%%%%%%%%%%%%%%%%%%%%%%%%%%%%%%%%%%%%%%% Fig_3 

%%%%%%%%%%%%%%%%%%%%%%%%%%%%%%%%%%%%%%%%%%%%%%%%%%%%%%%%%%%%% Fig_4
\begin{figure}[h]
\includegraphics[width=1.0\linewidth] {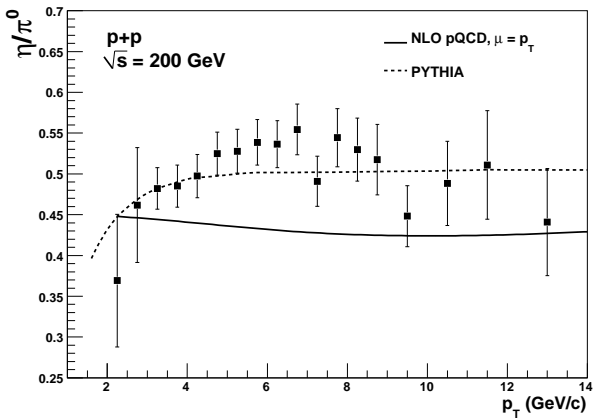}
\caption{
Cross section ratio for the midrapidity production of inclusive 
$\eta$ to $\pi^0$ mesons at $\sqrt{s}=200$~GeV as a function of $p_T$. 
The error bars show the statistical and systematic uncertainties added 
in quadrature. The solid curve shows the ratio of the NLO pQCD 
calculations shown in Fig.~\ref{subprocess_frac} and the corresponding 
one for the $\pi^0$. The dashed curve shows the result of a {\sc pythia} 
Monte-Carlo simulation. See text for details.}
\label{eta_cs_ratio}
\end{figure}

%==============================================  Table_II
\begin{table}[b]
 \caption{\label{table:ratio}
Results, statistical, and systematic (type-$B_1$, type-$B_2$) 
uncertainties of the measured $\eta$ to $\pi^0$ cross section ratio 
from the 2006 data set.}
 \begin{ruledtabular}
 \begin{tabular}{ccccc}
  $p_T$ (GeV/$c$) & $\eta/\pi^0$&  Stat. unc.  & type-$B_1$ & type-$B_2$ \\\hline
2.25 &  0.369 & 0.020 & 0.078 & 0.007 \\
2.75 &  0.462 & 0.024 & 0.065 & 0.009 \\
3.25 &  0.482 & 0.002 & 0.022 & 0.013 \\
3.75 &  0.485 & 0.003 & 0.022 & 0.014 \\
4.25 &  0.497 & 0.004 & 0.022 & 0.014 \\
4.75 &  0.525 & 0.005 & 0.023 & 0.010 \\
5.25 &  0.528 & 0.007 & 0.024 & 0.011 \\
5.75 &  0.539 & 0.009 & 0.024 & 0.011 \\
6.25 &  0.536 & 0.012 & 0.024 & 0.011 \\
6.75 &  0.555 & 0.015 & 0.025 & 0.011 \\
7.25 &  0.491 & 0.019 & 0.022 & 0.010 \\
7.75 &  0.544 & 0.024 & 0.024 & 0.011 \\
8.25 &  0.530 & 0.029 & 0.024 & 0.011 \\
8.75 &  0.518 & 0.035 & 0.023 & 0.010 \\
9.5 &  0.448 & 0.030 & 0.020 & 0.009 \\
10.5 &  0.488 & 0.045 & 0.022 & 0.014 \\
11.5 &  0.511 & 0.058 & 0.023 & 0.023 \\
13.0 &  0.441 & 0.058 & 0.020 & 0.024 \\
 \end{tabular}
 \end{ruledtabular}
 \end{table}

Constraining the fragmentation function further should be possible by 
including precise $\eta$ to $\pi^0$ cross section ratios in the 
extraction. The cross section ratio, as a function of $p_T$, is given 
in Fig.~\ref{eta_cs_ratio} and Table~\ref{table:ratio}. The ratio has 
been extracted in a single pass over the same data set, thus 
minimizing systematic uncertainties. In particular, the large 
normalization uncertainty of 9.7\% arising from the luminosity 
calculation cancels completely. Also all other systematic 
uncertainties are assumed to either cancel or be reduced to a 
negligible amount with the exception of the following. The systematic 
uncertainties due to the $\eta$ and $\pi^0$ peak extraction 
(type-$B_1$) and due to the correction for possible merging of the two 
$\pi^0$ decay photons into a single cluster (type-$B_2$) do not 
cancel. Furthermore, while the uncertainties on the high-$p_T$ trigger 
efficiency (type-$B_2$) are assumed to cancel above $p_T = 
4.5\,$GeV/$c$ where the efficiency is flat, a remaining 2\% 
uncertainty on the ratio is assigned for differences in the trigger 
turn-on curve for $3 < p_T < 4.5\,$GeV/$c$. Finally, the systematic 
uncertainty on the acceptance (type-$B_2$) is reduced to a $p_T$ 
independent contribution of 2\%. The ratio is presented up to $p_T = 
14$~GeV/$c$ only, as beyond this point the statistical and systematic 
uncertainties become rather large. The latter is due to the fact that 
for increasing transverse momenta the two photons from the $\pi^0$ 
have a strongly increasing probability of being reconstructed as only 
a single cluster in the calorimeter, leading to a rather large 
systematic uncertainty arising from the correction for this effect.

Except for an initial increase due to the different meson masses, the 
data do not exhibit a strong dependence on $p_T$. This hints towards a 
similar dependence of the $\eta$ and $\pi^0$ fragmentation functions 
on the energy fraction of the parton carried by the hadron. Fitting 
the cross section ratio, including its statistical and type-$B_1$ 
systematic uncertainty, to a constant, gives $R_{\eta / \pi^0} = 0.51 
\pm 0.01$ ($\chi^2$/ndf = 18.3/17), with a remaining systematic 
uncertainty of 0.01 from the type-$B_2$ systematic uncertainty. This 
result does not change when fitting the data above $p_T = 3$~GeV/$c$ 
($\chi^2$/ndf = 14.9/15) instead of fitting the full range.

Within the uncertainties, the present measurement of $\eta/\pi^0$ is 
consistent with all previous measurements in $p+p$ collisions, going 
back to the measurement reported in Ref.~\cite{Busser:1974yj}. A 
detailed comparison of subsequent measurements is summarized in 
\cite{Adler:2006bv}. The observed ratio is in good agreement with a 
{\sc pythia} 6.131~\cite{Sjostrand:2000wi} calculation~\cite{Adler:2006bv} 
shown in the same figure, which is using the default settings and the 
Lund string fragmentation model~\cite{Andersson:1983ia}. The solid 
line in Fig.~\ref{eta_cs_ratio} shows the ratio of the NLO pQCD 
calculations at a scale $\mu = p_T$ (see Fig.~\ref{eta_cs}) and the 
corresponding one for the $\pi^0$ using the same PDF but the $\pi^0$ 
fragmentation function of Ref.~\cite{deFlorian:2007aj}. Note that the 
shape of this calculated cross section ratio is not necessarily well 
determined as the statistical uncertainty on the $\eta$ fragmentation 
function, defined by $\Delta \chi^2 = 2$\%, results in an uncertainty 
on the $\eta$ cross section between about 5\% and 9\%, depending on 
$p_T$~\cite{Aidala:2010bn}.

The calculated ratio underestimates the data even though the $\eta$ 
cross section presented in this paper and earlier PHENIX $\pi^0$ data 
are part of the input in the extraction of the fragmentation 
functions. This indicates that the constraints from the separate fits 
are less stringent than fitting the cross section ratio directly. This 
is obvious from the fact that some of the experimental systematic 
uncertainties cancel in the ratio as already discussed above, in 
particular the overall normalization uncertainty of 9.7\% due to the 
uncertainty in the luminosity measurement. For example, the earlier 
PHENIX data used in the extraction of the $\pi^0$ fragmentation 
functions was scaled by a factor of 1.09~\cite{deFlorian:2007aj} in 
the fit, which is within the experimental normalization uncertainty, 
but leads to a smaller calculated cross section ratio as can be seen 
in Fig.~\ref{eta_cs_ratio}. Also, the dependence of the calculated 
$\eta$ and $\pi^0$ cross sections on the theoretical scale, as shown 
in, e.g., Fig.~\ref{eta_cs}, largely cancels in the calculation of the 
ratio~\cite{Indumathi:2008mu}. Hence it appears that improved 
constraints on $\eta$ and $\pi^0$ fragmentation functions can be 
derived by directly including the data on the $\eta$/$\pi^0$ cross 
section ratio in the fit.

\section{Double Helicity Asymmetry for $\eta$ Mesons}

Experimentally, the double helicity asymmetry 
(Eq.~\ref{cross_sec_asy}) translates into
\begin{equation} %\nonumber
A_{LL} = \frac{1}{|P_B||P_Y|}\frac{N_{++}-RN_{+-}}{N_{++}+RN_{+-}}, \quad
{\rm with} \quad
R\equiv\frac{L_{++}}{L_{+-}},
\end{equation}
where $N_{++}$ ($N_{+-}$) is the experimental yield for the case where 
the beams have the same (opposite) helicity. The polarizations of the 
two colliding beams at RHIC are denoted by $P_B$ and $P_Y$. The 
relative luminosity $R$ is measured by a coincident signal in the two 
BBCs that satisfies the vertex cut. Uncertainties on $A_{LL}$ of $2 
\times 10^{-4}$ ($7 \times 10^{-4}$) for the 2005 (2006) data due to 
relative luminosity uncertainties are uncorrelated between years.  
The asymmetries and uncertainties are combined by weighting by all 
year-to-year uncorrelated uncertainties in each $p_T$ bin. The results 
are given in Table~\ref{table:asym}.

The degree of polarization is determined from the combined information 
of a polarized-proton on carbon ($\vec {p}C$) 
polarimeter~\cite{Nakagawa:2008zzb}, using an unpolarized ultra--thin 
carbon ribbon target, and from elastic $\vec {p}+\vec{p}$ scattering, 
using a polarized atomic hydrogen gas-jet target~\cite{Okada:2005gu}. 
The average polarization value for the data from 2005 (2006) is 0.49 
(0.57).  There is a relative uncertainty of 4.8\% in the product of 
the beam polarizations, correlated between the 2005 and 2006 data 
sets, which is a scale uncertainty on the combined asymmetry result, 
affecting both the central values and the statistical uncertainties 
such that the statistical significance of the measurement from zero is 
preserved. Uncertainties on the products of the beam polarizations 
that are uncorrelated between years are combined using the same weight 
factors as for the uncertainties due to relative luminosity, and given 
in Table~\ref{table:asym}. In order to avoid false asymmetries due to 
a possible variation of detector response versus time or due to a 
possible correlation of detector performance with the RHIC bunch 
structure, all four helicity combinations in the colliding bunches are 
present within four consecutive bunch crossings. Possible transverse 
components of the beam polarizations at the PHENIX interaction point 
are monitored by measuring the spin dependence of very forward neutron 
production \cite{Bazilevsky:2006vd} in the zero degree calorimeters 
\cite{Adler:2000bd}.  The average transverse component of the product 
in the 2005 data set is less than $0.014 \pm 0.003$, described in more 
detail in~\cite{Adare:2007dg}, and was measured to be negligible in 
the 2006 data set.

The double helicity asymmetry for $\eta$ production,
\begin{equation} \nonumber
A_{LL}^{\eta} = \frac{A_{LL}^{\eta+BG}-rA_{LL}^{BG}}{1-r}, \quad \text {with}  \quad
r\equiv\frac{N^{BG}}{N^{\eta}+N^{BG}},
\end{equation}
can be calculated by separately measuring the asymmetry in the 
$2\sigma$ window around the mean of the $\eta$ peak 
($A_{LL}^{\eta+BG}$) and in the sidebands ($A_{LL}^{BG}$) defined 
above.

%%%%%%%%%%%%%%%%%%%%%%%%%%%%%%%%%%%%%%%%%%%%%%%%%%%%%%%%%%%%% Fig_5
\begin{figure}[tb]
\includegraphics[width=1.0\linewidth]{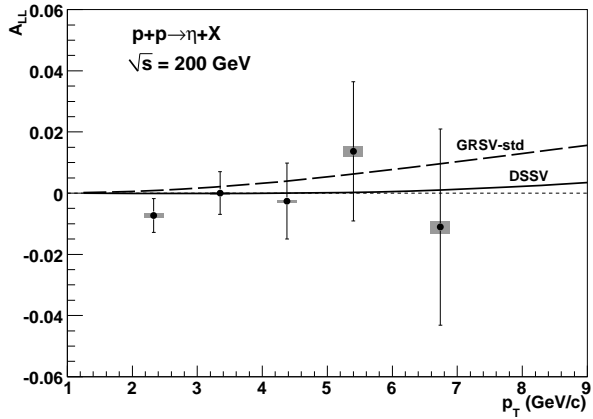}
\caption{
Double helicity asymmetry for midrapidity inclusive $\eta$ production 
from the combined 2005 and 2006 data at $\sqrt{s}=200$~GeV as a 
function of $p_T$.  The grey boxes are point-to-point systematic 
uncertainties due to polarization and relative luminosity 
uncertainties and are correlated point-to-point, moving all points in 
the same direction but not by the same factor. An additional 
systematic uncertainty of 4.8\% on the vertical scale due to the 
uncertainty in the beam polarizations is not shown. The results are 
compared to NLO pQCD calculations using two different sets of 
polarized PDFs\cite{Gluck:2000dy,deFlorian:2008mr}. See text for 
details.}
\label{asys}
\end{figure}

The latter is consistent with zero. The resulting background corrected 
asymmetry for $\eta$ production as a function of $p_T$ from the 
combined 2005 and 2006 data is shown in Fig.~\ref{asys} and tabulated 
in Table~\ref{table:asym}. It is consistent with zero over the 
measured range as can be expected based on the similar contributions 
of the various scattering subprocesses to the $\eta$ and $\pi^0$ 
production shown in Fig.~\ref{subprocess_frac} and the fact that the 
double helicity asymmetry for $\pi^0$ production \cite{Adare:2008px} 
is consistent with zero as well.

As can be seen in Fig.~\ref{asys}, the $\eta$ double helicity 
asymmetry is in agreement with NLO pQCD calculations using the above 
mentioned fragmentation functions and two different sets of polarized 
PDFs\cite{Gluck:2000dy,deFlorian:2008mr} as an input to the code of 
Ref.~\cite{Jager:2002xm}.

These data can be used in global fits in order to further constrain 
polarized PDFs, in particular the helicity--dependent gluon PDF. In 
the future, with improved statistics and the availability of 
flavor-separated fragmentation functions, double-helicity asymmetries 
in $\eta$ production can potentially constrain the polarized strange 
quark PDF ($\Delta s$) due to the additional $s$-quark contribution in 
the $\eta$ wave function. Special interest in $\Delta s$ arises from 
the fact that its value is negative, when extracted from analyses of 
inclusive DIS data, using hyperon decay data and assuming SU(3) flavor 
symmetry~\cite{Airapetian:2007mh,Alexakhin:2006vx}, but consistent 
with zero, when directly extracted from semi-inclusive DIS data 
\cite{Airapetian:2004zf,Airapetian:2008qf,Collaboration:2010ub}. 
Global fits can constrain PDFs by simultaneously describing a wide 
variety of experimental channels over a range of kinematics with 
different sensitivities, different experimental systematic 
uncertainties, and different sources of theoretical uncertainty. Thus, 
the data presented here open up a valuable new channel to improve 
knowledge of polarized PDFs.

\section{Summary}

The double helicity asymmetry in $\eta$ production is measured and 
found to be consistent with zero in the transverse momentum range 
between 2~GeV/$c$ and 9~GeV/$c$. The $\eta$ cross section is 
determined over seven orders of magnitude between 2~GeV/$c$ and 
20~GeV/$c$ in transverse momentum. In particular due to the wide range 
in transverse momentum these data serve as important input for the 
extraction of fragmentation functions. The $\eta$ to $\pi^0$ cross 
section ratio as a function of $p_T$ has been extracted in a single 
pass over the same data set, thus minimizing systematic uncertainties. 
A fit to a constant above $p_T = 2$~GeV/$c$ or $p_T = 3$~GeV/$c$ 
yields a value of 
$R_{\eta / \pi^0} = 0.51 \, \pm \, 0.01^{\rm stat}\, \pm \, 0.01^{\rm syst}$. 
The inclusion of these data on the ratio in 
future fragmentation function extractions should allow for more 
precise results for both particle species. This opens up the 
possibility to include the data on the double helicity asymmetry in 
future NLO pQCD fits in order to further constrain the polarized 
parton distribution functions, in particular the helicity--dependent 
gluon distribution function.

%==============================================  Table_III
 \begin{table}[t]
 \caption{\label{table:asym}
Double-helicity asymmetry values and uncertainties for combined 2005 
and 2006 data sets.  Systematic uncertainties given are type $B_2$, 
scaling all points in the same direction but not by the same factor, 
and are due to polarization ($\sigma_{sys}^P$) and relative luminosity 
($\sigma_{sys}^R$) uncertainties that are uncorrelated between years.  
There is an additional type $C$ systematic uncertainty of 4.8\% on the 
vertical scale due to uncertainty in the beam polarizations that is 
correlated between years. }
 \begin{ruledtabular}
 \begin{tabular}{ccccc}
  \meanpt (GeV/$c$) & $A_{LL}$ &  $\sigma_{\rm stat}$  & $\sigma_{\rm syst}^P$ 
                    &  $\sigma_{\rm syst}^R$\\\hline
   2.33  &  -0.0073  &  0.0055    &  0.0006 & 0.0004 \\
   3.35  &  0.0000   &  0.0070    &  0.0000 & 0.0004 \\
   4.38 &  -0.0026   &  0.0124    &  0.0002 & 0.0004 \\
   5.40 &   0.0137   &  0.0228    &  0.0016 & 0.0004 \\
   6.74 &  -0.0111   &  0.0320    &  0.0021 & 0.0004 \\
 \end{tabular}
 \end{ruledtabular}
 \end{table}

%%%%%%%%%%%%%%%%%%%%%%%%%  Acknowledgements

\begin{acknowledgments}

%\section{Acknowledgements}   % Run-6 long from for PRC, PLB, etc.

We thank the staff of the Collider-Accelerator and Physics
Departments at Brookhaven National Laboratory and the staff of
the other PHENIX participating institutions for their vital
contributions.  
We also thank M. Stratmann and R. Sassot for fruitful discussions.
We acknowledge support from the Office of Nuclear Physics in the
Office of Science of the Department of Energy,
the National Science Foundation, 
a sponsored research grant from Renaissance Technologies LLC, 
Abilene Christian University Research Council, 
Research Foundation of SUNY, 
and Dean of the College of Arts and Sciences, Vanderbilt University 
(USA),
Ministry of Education, Culture, Sports, Science, and Technology
and the Japan Society for the Promotion of Science (Japan),
Conselho Nacional de Desenvolvimento Cient\'{\i}fico e
Tecnol{\'o}gico and Funda\c c{\~a}o de Amparo {\`a} Pesquisa do
Estado de S{\~a}o Paulo (Brazil),
Natural Science Foundation of China (People's Republic of China),
Ministry of Education, Youth and Sports (Czech Republic),
Centre National de la Recherche Scientifique, Commissariat
{\`a} l'{\'E}nergie Atomique, and Institut National de Physique
Nucl{\'e}aire et de Physique des Particules (France),
Ministry of Industry, Science and Tekhnologies,
Bundesministerium f\"ur Bildung und Forschung, Deutscher
Akademischer Austausch Dienst, and Alexander von Humboldt Stiftung (Germany),
Hungarian National Science Fund, OTKA (Hungary), 
Department of Atomic Energy (India), 
Israel Science Foundation (Israel), 
National Research Foundation and WCU program of the 
Ministry Education Science and Technology (Korea),
Ministry of Education and Science, Russia Academy of Sciences,
Federal Agency of Atomic Energy (Russia),
VR and the Wallenberg Foundation (Sweden), 
the U.S. Civilian Research and Development Foundation for the
Independent States of the Former Soviet Union, 
the US-Hungarian Fulbright Foundation for Educational Exchange,
and the US-Israel Binational Science Foundation.

\end{acknowledgments}

%\bibliography{ppg107x0}

\end{document}